\documentstyle[12pt,epsf]{ioplppt}

\def\bare{\mathaccent"7017 }

\begin{document}

\jl{1}

\title{Equation of state for directed percolation}
\author{H K Janssen, \"U Kutbay and K Oerding}

\address{Institut f\"ur Theoretische Physik III,\\
Heinrich-Heine-Universit\"at, 40225 D\"usseldorf, Germany}

\begin{abstract}
Using field-theoretic renormalization group methods we calculate
the equation of state for non-equilibrium systems belonging to
the universality class of directed percolation (Gribov process) to
second order in $\epsilon = 4-d$. By introducing a parametric
representation the result can be written to this order in a very
simple form. We use our result to obtain a universal
amplitude ratio to second order in $\epsilon$.
\end{abstract}

\maketitle

\section{Introduction}

Numerous physical systems that show continuous non-equilibrium phase
transitions display near the transition point space-time
structures that can be described by directed percolation clusters.
Well-known examples are epidemic processes without
immunization~\cite{Murray}, poisoning of catalytic surfaces~\cite{DB88},
roughening transitions in growth processes~\cite{KW89}, Schl\"ogl's
reaction models~\cite{Schlogl,J81}, contact
processes~\cite{Harris,Liggett} and certain cellular
automata~\cite{Kinzel83,Kinzel85}. The model known in particle physics
as reggeon field theory~\cite{Grib67,Grib68,Moshe} has also been
related to directed percolation and epidemic
processes~\cite{GS78,GT79,CS80,Obuk}. All these systems share the
property that their critical behaviour is characterized by the existence
of an absorbing state. This property determines their universality
class~\cite{J81,Grass82}.

The behaviour of these systems near the transition point is 
characterized by universal quantities such as critical exponents.
The exponents for directed percolation have been calculated to second
order in $\epsilon=4-d$~\cite{J81} and---numerically---for $d=1$ and
$d=2$~(for recent results see, e.g., \cite{GZ96,Jensen96,LSMJ}).
It is well-known from the theory of equilibrium critical phenomena
that universality not only holds for critical exponents but also for
certain amplitude ratios and scaling functions~\cite{BLZ,Wallace,PHA}.
An example for a universal relationship between physical quantities
near a critical point is the equation of state. In the present paper
we derive an equation that describes the stationary state of epidemic
processes with an absorbing state near the transition point.

Before we turn to the mesoscopic description used in the analysis
presented below let us briefly discuss two examples for microscopic models
that belong to the universality class of directed percolation.
Schl\"ogl's autocatalytic reaction scheme~\cite{Schlogl} is defined by
the reaction equations
\begin{equation}
X + A \left.\begin{array}{c} \mbox{\small $k_{1}$}\\
\longrightarrow \\ \longleftarrow \\ \mbox{\small $k_{1}^{\prime}$}
\end{array}\right. 2 X , \qquad
X \left.\begin{array}{c} \mbox{\small $k_{2}$}\\
\longrightarrow \\ \longleftarrow \\ \mbox{\small $k_{3}$}
\end{array}\right. B  \label{schl}
\end{equation}
where the concentration of $A$'s and $B$'s is kept constant (e.g.,
by reservoirs). Here two remarks are in order: (i) If $A=B$ (i.e., the
particles are of the same type) and their concentration is not kept
constant (but governed by the reaction-diffusion dynamics) the system
belongs to a different universality class. In that case the total
particle density is conserved and affects the long-time dynamics of
the system~\cite{KSS,WOH}. (ii) Only if the reaction $B \rightarrow X$
has zero probability the system has an absorbing state.
In the case of chemical systems
it is natural to take this process into account since back reactions
can in general not be avoided. Thus the B-particles represent a source
for X. Below a critical value of the reaction rate $k_{2}$ the density of
$X$-particles is nonvanishing in the stationary state for $k_{3} \to 0$.

In the theory of infectious diseases~\cite{Murray} one considers
two types of individuals: Susceptibles S who can catch the
disease and infectives I who have the disease and can transmit it.
The dynamics of the system is characterized by the following
processes:
\begin{enumerate}
\item Healthy (susceptible) individuals may catch the disease by contact
with infectives. Infected individuals may recover spontaneously and
thereby become susceptible again.
\item Interaction of infected individuals can lead to recovery
(``saturation''). 
\item Diffusive spreading of the disease.
\item Susceptibles may catch the disease spontaneously.
\end{enumerate}
The process (iv) models for instance a continuos flow of germs into
the system. Note that the system has an absorbing state if (and only if)
the process (iv) has zero probability. 

A mesoscopic description for the dynamics near the transition point
requires a single density $n({\bf r}, t)$ which represents the slow
degrees of freedom. In the case of Schl\"ogl's model~(\ref{schl})
$n$ corresponds to the density of $X$-particles, in the context of
poisoning of a catalytic surface $n$ is the density of vacant sites and
in epidemic processes it represents the density of infectives.
The density satisfies the Langevin equation
\begin{equation}
\partial_{t} n = \lambda \nabla^{2} n + R[n] n + \lambda q + \zeta
\label{lang}
\end{equation}
where the reaction rate $R[n]$ can be expanded in powers of $n$.
In order to investigate the universal properties near the transition
point we only need the two leading terms,
\begin{equation}
R[n] = \lambda (\tau + \frac{g}{2} n)
\end{equation}
where $g>0$ and $\tau$ is a temperature-like critical parameter.
For $q=0$ the system has an absorbing state with $n=0$ if the
random force $\zeta$ is multiplicative:
\begin{equation}
\langle \zeta({\bf r}, t) \zeta({\bf r}^{\prime}, t^{\prime})
\rangle = \lambda \tilde{g} n({\bf r}, t) \delta({\bf r}-
{\bf r}^{\prime}) \delta(t-t^{\prime})
\end{equation}
where we have again neglected irrelevant higher powers in $n$.
In the case of the reaction-diffusion system~(\ref{schl}) the
(constant) source term $q$ vanishes for a zero reaction rate $k_{3}$.
By a simple rescaling it is possible to render the coupling constants
$g$ and $\tilde{g}$ equal.

\section{Renormalized field theory and perturbational calculation
of the equation of state}

Hereafter we work with the dynamic functional~\cite{J81,J76,DeD,BJW,J92}
\begin{equation}
{\cal J}[\tilde{s}, s] = \int \d t \int \d^{d}r \tilde{s} \Big[
\partial_{t} s + \lambda ( \tau - \nabla^{2}) s + \frac{\lambda g}{2}
s(s-\tilde{s}) - \lambda q \Big] \label{J}
\end{equation}
where $s \propto n$ and $\tilde{s}$ denotes a Martin-Siggia-Rose
response  field~\cite{MSR}. Response and correlation functions can be
computed by integrating products of the fields $\tilde{s}$ and $s$ with
the weight $\exp(-{\cal J})$. We are especially interested in the
average $M = \langle s \rangle$ which corresponds to the mean particle
density. The external field $\lambda q$ describes the spontaneous
creation of particles and breaks the symmetry $\tilde{s}(t)
\leftrightarrow -s(-t)$.  

In order to obtain the equation of state---i.e., $M$ as a function
of $\tau$ and $q$ in the stationary state---we perform the shift
$s = M + \phi$ und determine $M$ by the no-tadpole requirement
$\langle \phi \rangle =0$. Since the stationary state is homogeneous
the shift leads to the functional
\begin{equation}
{\cal J}_{\phi}[\tilde{\phi}, \phi] = {\cal J}_{0}[\tilde{\phi}, \phi]
+ {\cal J}_{G}[\tilde{\phi}, \phi] + {\cal J}_{I}[\tilde{\phi}, \phi]
\label{Jphi}
\end{equation}
where $\tilde{\phi} = \tilde{s}$,
\begin{equation}
{\cal J}_{0}[\tilde{\phi}, \phi] = \int \d t \int \d^{d}r \lambda
\tilde{\phi} \Big[M (\tau M + \frac{g}{2} M) - q\Big]
\end{equation}
\begin{equation}
{\cal J}_{G}[\tilde{\phi}, \phi] = \int \d t \int \d^{d}r \tilde{\phi}
\Big[ \partial_{t}\phi + \lambda (\tau + gM - \nabla^{2}) \phi -
\frac{\lambda g}{2} M \tilde{\phi} \Big] \label{Jgauss}
\end{equation}
and
\begin{equation}
{\cal J}_{I} = \int \d t \int \d^{d}r \frac{\lambda g}{2}\tilde{\phi}
\phi (\phi - \tilde{\phi}) . \label{Jint}
\end{equation}
The equation of state can now be written in the form
\begin{equation}
\lambda q = \lambda M ( \tau + \frac{g}{2} M ) -
\begin{picture}(45,10)
\put(0,-13){\epsfxsize=15mm\epsfbox{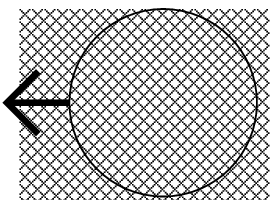}}
\end{picture}
\end{equation}
where the bubble represents the sum of all one-particle irreducible
Feynman diagrams of the field theory~(\ref{Jphi})-(\ref{Jint}) with
one external $\tilde{\phi}$-leg. Details of the diagrammatic analysis
are given in the appendix. After dimensional regularization of
ultraviolet divergences one obtains at two-loop order
\begin{equation}
\fl q =  M \left[\tau + \frac{g}{2} M - \frac{G_{\epsilon} g^{2}
\bar{\tau}^{1-\epsilon/2}}{ \epsilon (2-\epsilon)} -
\frac{g^{4}\bar{\tau}^{-\epsilon}}{24} \left( g M
\frac{1-\epsilon}{2} I_{1} + (2 I_{1}+3 I_{2}) \bar{\tau} \right)
\right] \label{perturb}
\end{equation}
where $G_{\epsilon} = \Gamma(1+\epsilon/2)/(4\pi)^{d/2}$
is a geometrical factor and
\begin{equation}
\bar{\tau} = \tau + g M .
\end{equation}
The integrals $I_{1}$ and $I_{2}$ are defined in the appendix.

The poles in $\epsilon$ can be absorbed into renormalizations of the
fields and the coupling constants. The renormalizations read
\begin{eqnarray}
\tilde{\phi} \to \bare{\tilde{\phi}} = Z^{1/2} \tilde{\phi} \qquad
\phi \to \bare{\phi} = Z^{1/2} \phi \qquad M \to \bare{M} = Z^{1/2}
M \nonumber \\
q \to \bare{q} = Z_{\lambda}^{-1} Z^{1/2} q \qquad \lambda \to
\bare{\lambda} = Z^{-1} Z_{\lambda} \lambda \qquad \tau \to \bare{\tau}
= Z_{\lambda}^{-1} Z_{\tau} \tau \\
g \to \bare{g} = Z^{-1/2} Z_{\lambda}^{-1} Z_{g} g \qquad
G_{\epsilon} g^{2} = u \mu^{\epsilon}
\nonumber 
\end{eqnarray}
where $\mu$ denotes an external momentum scale. The required
renormalization factors have been determined to two loop order in
Ref.~\cite{J81}. The Z-factors are
\begin{eqnarray}
Z = 1 + \frac{u}{4 \epsilon}  + \left(\frac{7}{\epsilon}
- 3 + \frac{9}{2} \ln \frac{4}{3} \right) \frac{u^{2}}{32 \epsilon}
\nonumber
\\ Z_{\lambda} = 1 + \frac{u}{8\epsilon} + \left(\frac{13}{\epsilon}
- \frac{31}{4} + \frac{35}{2} \ln \frac{4}{3} \right)
\frac{u^{2}}{128 \epsilon} \\
Z_{\tau} = 1 + \frac{u}{2\epsilon} + \left(\frac{1}{2\epsilon^{2}}
- \frac{5}{32 \epsilon}\right) u^{2} \qquad
Z_{g} = 1 + \frac{u}{\epsilon} + \left( \frac{5}{4\epsilon^{2}}
- \frac{7}{16 \epsilon} \right) u^{2} . \nonumber
\end{eqnarray}

\section{Scaling form of the equation of state}

In order to improve the perturbative result~(\ref{perturb}) by the
renormalization group we exploit as usual the invariance of the
bare field theory with respect the a variation of $\mu$ at fixed
bare parameters, i.e.,
\begin{equation}
\left. \mu \frac{\d}{\d \mu} \right|_{0} \bare{q}(\bare{\tau},
\bare{M}, \bare{g}) = 0 .
\end{equation}
This equation---expressed by renormalized quantities---is the
renormalization group equation
\begin{equation}
\Big[\mu \partial_{\mu} + \kappa \tau \partial_{\tau} - \frac{1}{2}
\gamma M \partial_{M} + \beta \partial_{u} + \zeta - \frac{1}{2}
\gamma \Big] q(\tau, M, u; \mu) = 0 \label{RGE}
\end{equation}
with the Wilson functions (to two-loop order)
\begin{eqnarray}
\gamma(u) = \left.\frac{\d \ln Z}{\d \ln \mu}\right|_{0} = -\frac{u}{4}
+ \left(3 - \frac{9}{2} \ln \frac{4}{3}\right) \frac{u^{2}}{16}
\nonumber \\
\kappa(u) = \left.\frac{\d \ln \tau}{\d \ln \mu} \right|_{0} =
\frac{3 u}{8} - \left( \frac{49}{4} + \frac{35}{2} \ln \frac{4}{3} 
\right) \frac{u^{2}}{64} \nonumber \\
\zeta(u) = \left.\frac{\d \ln \lambda}{\d \ln \mu}\right|_{0} =
- \frac{u}{8} + \left(\frac{17}{2} - \ln\frac{4}{3} \right)
\frac{u^{2}}{128}
\\
\beta(u) = \left.\frac{\d u}{\d \ln \mu}\right|_{0} = u \left[-\epsilon
+ \frac{3 u}{2} - \left( 169 + 106 \ln \frac{4}{3} \right)
\frac{u^{2}}{128} \right] . \nonumber
\end{eqnarray}
The renormalization group equation~(\ref{RGE}) can be solved by
characteristics. At the fixed point ($u_{\star}$ with
$\beta(u_{\star})=0$)
\begin{equation}
u_{\star} = \frac{2\epsilon}{3} \left[ 1 + \left(\frac{169}{192}
+ \frac{53}{96} \ln \frac{4}{3} \right) \frac{2\epsilon}{3} 
+ \Or(\epsilon^{2}) \right]
\end{equation}
the result
(combined with dimensional analysis) reads
\begin{equation}
q(\tau, M, u; \mu) = \mu^{2+d/2} l^{z + (d-\eta)/2}
q\Big(\mu^{-2} l^{-1/\nu} \tau, \mu^{-d/2} l^{-(d+\eta)/2} M,
u_{\star}; 1\Big) \label{flow}
\end{equation}
with the critical exponents~\cite{J81}
\begin{eqnarray}
\eta = \gamma(u_{\star}) = -\frac{\epsilon}{6} \left[
1 + \left( \frac{25}{288} + \frac{161}{144} \ln \frac{4}{3}
\right) \epsilon + \Or(\epsilon^{2}) \right] \nonumber \\
\nu = \frac{1}{2-\kappa(u_{\star})} = \frac{1}{2} +
\frac{\epsilon}{16} \left[1 + \left( \frac{107}{288} -
\frac{17}{144} \ln \frac{4}{3} \right) \epsilon +
\Or(\epsilon^{2}) \right] \nonumber \\
z = 2 + \zeta(u_{\star}) = 2 - \frac{\epsilon}{12} \left[
1 + \left( \frac{67}{288} + \frac{59}{144} \ln\frac{4}{3} 
\right) \epsilon + \Or(\epsilon^{2}) \right] \\
\beta = \frac{\nu (d+\eta)}{2} = 1 - \frac{\epsilon}{6} \left[
1 - \left( \frac{11}{288} - \frac{53}{144} \ln \frac{4}{3} \right)
\epsilon + \Or(\epsilon^{2}) \right] \nonumber \\
\delta = 1 + \frac{\nu (z-\eta)}{\beta} = 2 + \frac{\epsilon}{3}
\left[ 1 + \left( \frac{85}{288} + \frac{53}{144} \ln \frac{4}{3}
\right) \epsilon + \Or(\epsilon^{2}) \right] . \nonumber
\end{eqnarray}
and the scaling function
\begin{eqnarray}
\fl q(\tau, M, u_{\star}, \mu) = M \left[ \tau + \frac{g}{2} M -
\frac{\epsilon}{6} \bar{\tau} \left( A - B \ln (\bar{\tau}/\mu^{2})
- \frac{\epsilon}{12} (\ln (\bar{\tau}/\mu^{2}))^{2} \right)
\right. \nonumber \\
+ \left. \frac{\epsilon^{2}}{72} g M \left( 1+I - 2 \ln
(\bar{\tau}/\mu^{2}) + (\ln (\bar{\tau}/\mu^{2}))^{2} \right)
\right] \label{pertren}
\end{eqnarray}
where
\begin{eqnarray}
A = 1 + \epsilon \left( \frac{13}{288} + \frac{53}{144}
\ln \frac{4}{3} - \frac{1}{4} \ln 3 - \frac{I}{3} \right) \\
B = 1 + \epsilon \left( \frac{37}{288} + \frac{53}{144}
\ln \frac{4}{3} \right).
\end{eqnarray}
An integral representation for $I = -2.3439072\ldots$ is
given in the appendix.
In order to simplify the writing we will use hereafter the
dimensionless quantities defined by
\begin{equation}
\mu^{-2} \tau \to \tau \qquad \mu^{-2} g M \to M \qquad
\mu^{-4} g q \to q .
\end{equation}

While equation~(\ref{pertren}) cannot be used directly to investigate
the equation of state in the critical region ($\tau$, $M \to 0$)
one may map the critical region by an appropriate choice of the flow
parameter $l$ in~(\ref{flow}) on scales on which perturbation theory
can be applied. For $l = M^{\nu/\beta}$ equations~(\ref{pertren})
and~(\ref{flow}) yield the scaling form
\begin{equation}
a q = M^{\delta} F(b \tau M^{-1/\beta})
\label{eqstate1} 
\end{equation}
with the scaling variables
$x = b \tau M^{-1/\beta}$ and $z = a q M^{-\delta}$
and the universal scaling function
\begin{eqnarray}
\fl F(x) = x+1 + \frac{\epsilon}{6} K \left[ (x+2) \ln (x+2)
- 2 (x+1) \ln 2 \right] \nonumber \\
+ \frac{\epsilon^{2}}{72} \left[ (x+4) (\ln(x+2))^{2} - 4 (x+1)
(\ln 2)^{2} \right] + \Or(\epsilon^{3}) \label{scalfun}
\end{eqnarray}
where
\begin{equation}
K = 1 + \epsilon \left( \frac{85}{288} + \frac{29}{72} \ln 2
- \frac{53}{144} \ln 3 \right) = 1 + \epsilon \ 0.1699728\ldots.
\end{equation}
The coefficients $a$ and $b$ in equation~(\ref{eqstate1}) are
\begin{eqnarray}
a = 2 \left[1 + \frac{\epsilon}{3} + \frac{\epsilon^{2}}{18}
\left( \frac{85}{48} + \frac{53}{12} \ln 2 - \frac{89}{24} \ln 3
- \frac{5}{2} I) + \Or(\epsilon^{3})
\right) \right] \label{a} \\
b = 2 \left[ 1 + \frac{1+\ln 2}{6} \epsilon +
\frac{\epsilon^{2}}{36} \left( \frac{13}{48} -
\frac{89}{24} \ln 3 - 3 I
\right.\right. \nonumber \\ \left. \left. \qquad \qquad
+ (\frac{99}{16} +  \frac{59}{12} \ln 2 - \frac{53}{24}
\ln 3 ) \ln 2 \right) + \Or(\epsilon^{3})
\right] .  \label{b}
\end{eqnarray}
They have been introduced in order to normalize the
scaling function, i.e., \begin{equation}
F(0) = 1 \qquad F(-1) = 0 .
\end{equation}

The equation of state~(\ref{eqstate1}) satisfies the correct scaling
behaviour in the critical region but the scaling function~(\ref{scalfun})
can only be used if $x$ is not too large. Since $q$ is an analytic
function of $M$ for positive $\tau$ (for equilibrium systems this was
first discussed by Griffiths~\cite{Griff}) the
approximation~(\ref{scalfun}) breaks down for $M \ll \tau^{\beta}$.
In order to study this limit one may choose the flow parameter
in~(\ref{flow}) as $l = \tau^{\nu}$. One arrives at
\begin{equation}
a q = (b \tau)^{\beta \delta} \bar{F}(M (b \tau)^{-\beta})
\end{equation}
with the scaling variables $y = M (b \tau)^{-\beta} = x^{-\beta}$
and $\bar{z} = a q (b\tau)^{-\beta \delta} = z y^{\delta}$ and
the scaling function
\begin{eqnarray}
\fl \bar{F}(y) = y \left\{1+y + \frac{\epsilon}{6} K \left[ (1+2y)
\ln (1+2y) - 2 (1+y) \ln 2 \right] \right. \nonumber \\
\left. + \frac{\epsilon^{2}}{72} \left[ (1+4y) (\ln(1+2y))^{2} - 4
(1+y) (\ln 2)^{2} \right] + \Or(\epsilon^{3}) \right\}
\label{scalfunbar}
\end{eqnarray}
which is fully analytic in $y$:
\begin{equation}
\bar{F}(y) = y \sum_{k=0}^{\infty} c_{k} y^{k} . \label{expan}
\end{equation}
Of course $y^{-\delta} \bar{F}(y)$ and $F(x)$ 
match in the $\epsilon$-expansion up to second order.

\section{Parametric form of the equation of state}

The full range of the variables $\tau$, $M$ (including the active
phase for $\tau < 0$) can be investigated by introducing a parametric
representation for the equation of state. For equilibrium systems
such a representation was first suggested by Schofield et
al.~\cite{SLH} and Josephson~\cite{Joseph} and later derived
in an $\epsilon$-expansion by Br\'ezin et al.~\cite{BWW72}.
To apply this method to directed percolation we write $\tau$ and
$M$ as
\begin{equation}
b \tau = R (1-\theta) \qquad M = R^{\beta} \theta/\theta_{0} .
\label{param}
\end{equation}
Inserting~(\ref{param}) with $\theta_{0}=2$ into~(\ref{eqstate1}) one
finds
\begin{equation}
a q = (R^{\beta}/2)^{\delta} H(\theta)
\label{eqstate3}
\end{equation}
with the simple scaling function
\begin{equation}
H(\theta) = \theta (2-\theta) + \Or(\epsilon^{3}) .\label{scalfun2}
\end{equation}

\begin{figure}[ht]
\caption{Sketch of the parametric representation~(\protect\ref{param}).
The straight lines correspond to a variation of $\theta$ at fixed $R$.}
\epsfxsize=500pt
\vspace*{-80mm}
\epsfbox{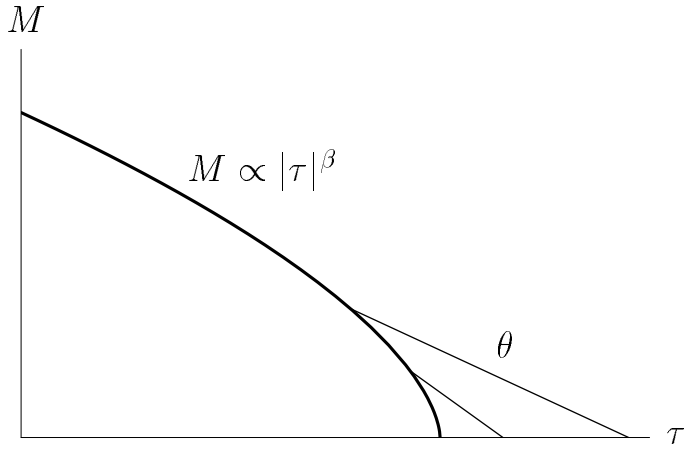}
\vspace*{-100mm}
\label{sketch}
\end{figure}

The parameter range $R\geq 0$, $0 \leq \theta < 2$ is required
to describe the whole phase diagram around the critical point.
The case $\tau < 0$, $q \to +0$ corresponds to the limit
$\theta \to 2$ (from below). (See figure~\ref{sketch}.)

As a simple application of the parametric representation
we briefly discuss the susceptibility
\begin{equation}
\chi = \left. \frac{\partial M}{\partial q}\right|_{\tau}
\end{equation}
which satisfies a power law for $q \to 0$,
\begin{equation}
\chi = \chi_{\pm} |\tau|^{-\gamma}
\end{equation}
where $\chi_{+}$ and $\chi_{-}$ correspond to the cases $\tau > 0$
and $\tau < 0$, respectively, and $\gamma = \beta (\delta-1)$.
To second order in $\epsilon$ the susceptibility can be written in
terms of the parameters~(\ref{param}) as
\begin{equation}
\chi = a 2^{\delta - 1} R^{-\gamma}
\frac{1-(1-\beta)\theta}{\beta\delta \theta (2-\theta) +
2 (1-\theta)^2} .
\end{equation}
Therefore the universal amplitude ratio $\chi_{-}/\chi_{+}$
can be expressed to this order by the exponent $\beta$ as
\begin{eqnarray}
\frac{\chi_{-}}{\chi_{+}} &= 2 \beta - 1 + O(\epsilon^{3}) =
1 - \frac{\epsilon}{3} \left[
1 - \left( \frac{11}{288} - \frac{53}{144} \ln \frac{4}{3} \right)
\epsilon + \Or(\epsilon^{2}) \right] \\
&= 1- \frac{\epsilon}{3}(1 + \epsilon \ 0.0676885\ldots ).
\end{eqnarray}

The parametric equation of state allows us also to express the
expansion coefficient $c_{0}$ in~(\ref{expan}) to second order
in $\epsilon$ by $\delta$. For this purpose we write
\begin{equation}
\fl x = 2^{1/\beta} \theta^{-1/\beta} (1-\theta) \qquad
y = \frac{\theta/2}{(1-\theta)^{\beta}} \qquad
z = \theta^{-\delta} H(\theta) \qquad \bar{z} =
\frac{2^{-\delta} H(\theta)}{(1-\theta)^{\beta \delta}}
\end{equation}
which gives to $\Or(\epsilon^{2})$
\begin{equation}
\frac{\bar{F}(y)}{y} = \frac{\bar{z}}{y} = 
\frac{2^{1-\delta}(2-\theta)}{(1-\theta)^{\beta (\delta-1)}} .
\end{equation}
For $y \sim \theta \to 0$ this becomes
\begin{eqnarray}
\fl c_{0} = 2^{2-\delta} + \Or(\epsilon^{3}) = 1 - \frac{\epsilon}{3}
\ln 2 - \frac{\epsilon^{2}}{864} \left( 85 + 164 \ln 2 - 106 \ln
3\right) \ln 2 + \Or(\epsilon^{3}) \\
= 1 - \epsilon \ 0.231049 - \epsilon^{2} \ 0.0659639 +
\Or(\epsilon^{3}) . \nonumber
\end{eqnarray}

\section{Conclusion}

We have derived a universal equation that describes the order
parameter of directed percolation systems as a function of the scaling
fields near the transition point to second order in $\epsilon=4-d$.
Using a parametric representation for the thermodynamic variables
our result could be written in a very simple form.
This analysis was motivated by the success of a parametric equation
of state [analogous to~(\ref{param})-(\ref{scalfun2})] for equilibrium
critical phenomena which was shown to be in good agreement with
experiments~\cite{SLH}.
Our hope is that our results will help to analyse data obtained by
Monte Carlo simulations. It would be especially interesting to see how
important the $O(\epsilon^{3})$ corrections to Eq.~(\ref{scalfun2})
are in practice.

\ackn

This work has been supported in part by the Sonderforschungsbereich
237 [Unordnung und Gro{\ss}e Fluktuationen (Disorder and Large
Fluctuations)] of the Deutsche Forschungsgemeinschaft.

\appendix

\section*{Appendix}

\begin{figure}[ht]
\caption{Contributions to the equation of state to two loop order.}
\epsfxsize=430pt
\vspace*{6mm}\hspace*{3mm}
\epsfbox{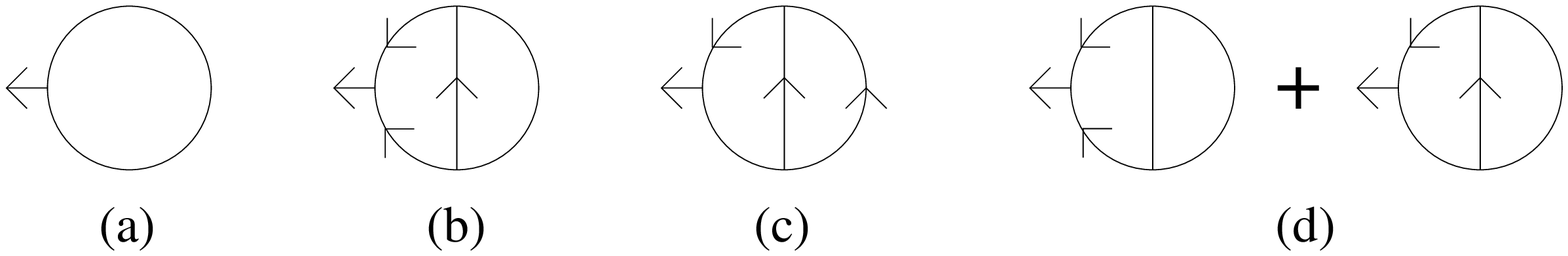}
\label{feynman}
\end{figure}

In this appendix we give some details of the diagrammatic analysis
required to compute the equation of state. The diagrams
contributing to two loop order are given in figure~\ref{feynman}.
The gaussian propagator follows from the action ${\cal J}_{G}$ in
equation~(\ref{Jgauss}) as
\begin{equation}
t \, \stackrel{{\bf q}}{\begin{picture}(45,15)
\put(0,-3){\epsfxsize=15mm\epsfbox{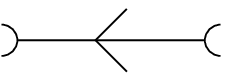}}
\end{picture}} t^{\prime} = \Theta(t-t^{\prime}) \exp\left(-\lambda
(\bar{\tau} + q^{2}) (t-t^{\prime}) \right) \label{prop}
\end{equation}
where $\Theta(t)$ denotes the step function and ${\bf q}$ is the
momentum carried by the line.
The correlator reads
\begin{equation}
t \, \stackrel{{\bf q}}{\begin{picture}(45,10)
\put(0,-1){\epsfxsize=15mm\epsfbox{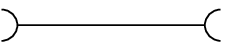}}
\end{picture}} t^{\prime} = \frac{g M}{2 (\bar{\tau} + q^{2})}
\exp\left(-\lambda (\bar{\tau} + q^{2}) |t-t^{\prime}| \right)
\end{equation}
and the vertices
\begin{equation}
\begin{picture}(20,20)
\put(0,-7){\epsfxsize=7mm\epsfbox{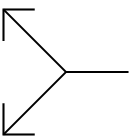}}
\end{picture} =
-\,\begin{picture}(20,20)
\put(0,-7){\epsfxsize=7mm\epsfbox{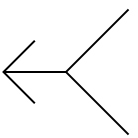}}
\end{picture} = \lambda g
\end{equation}
represent the interaction~(\ref{Jint}).

After integration over the internal time arguments the Feynman
diagrams are
\begin{eqnarray}
{\rm (a)} = - \frac{\lambda g^{2} M}{4} \int_{\bf p}
\frac{1}{\bar{\tau}+p^{2}} = \frac{G_{\epsilon}}{\epsilon
(2-\epsilon)} \lambda g^{2} M \bar{\tau}^{1-\epsilon/2}\\
{\rm (b)} = \frac{\lambda g^{4} M}{12} \int_{{\bf k}, {\bf p}}
\frac{1}{(\bar{\tau} + k^{2}) (\bar{\tau} + p^{2}) (\bar{\tau} +
({\bf k}+{\bf p})^{2})}
= \frac{\lambda g^{4} M}{12}  \bar{\tau}^{1-\epsilon} I_{1} \\
{\rm (c)} = \frac{\lambda g^{4} M}{8} \int_{{\bf k}, {\bf p}}
\frac{1}{(\bar{\tau}+k^{2})^{2} (3\bar{\tau} + k^{2} + p^{2} +
({\bf k} + {\bf p})^{2})} = \frac{\lambda g^{4} M}{8}
\bar{\tau}^{1-\epsilon} I_{2} \\
{\rm (d)} = -\frac{\lambda g^{5} M^{2}}{16} \int_{{\bf k}, {\bf p}}
\frac{1}{(\bar{\tau}+k^{2})^{2}(\bar{\tau}+p^{2})(\bar{\tau}+({\bf k}
+ {\bf p})^{2})} \\
= \frac{\lambda g^{5} M^{2}}{16} \frac{1}{3}
\frac{\partial}{\partial \bar{\tau}} \int_{{\bf k}, {\bf p}}
\frac{1}{(\bar{\tau}+k^{2})(\bar{\tau}+p^{2})(\bar{\tau}+({\bf k} +
{\bf p})^{2})} 
= \frac{\lambda g^{5} M^{2}}{16} \frac{1-\epsilon}{3}
\bar{\tau}^{-\epsilon} I_{1} \nonumber
\end{eqnarray}
where we have used the notation $\int_{{\bf k}} \cdots =
(2\pi)^{-d} \int \d^{d}k \cdots$.

The $\epsilon$-expansions of the integrals $I_{1}$ and $I_{2}$ read
\begin{eqnarray}
I_{1} &= \int_{{\bf k}, {\bf p}} \frac{1}{(1+k^{2}) (1+p^{2})
(1+({\bf k} + {\bf p})^{2})} \nonumber \\
&= -\frac{6 G_{\epsilon}^{2}}{\epsilon^{2}}
\left[ 1 + \frac{3 \epsilon}{2} + \frac{\epsilon^{2}}{4} (7+I)
+ O(\epsilon^{3}) \right]
\end{eqnarray}
with 
\begin{equation}
I = \int_{0}^{1} \d x \frac{\ln(x (1-x))}{1-x (1-x)} = 
-2.3439072\ldots .
\end{equation}
and
\begin{eqnarray}
I_{2} &= \int_{{\bf k}, {\bf p}} \frac{1}{(1+k^{2})^{2}
(3 + k^{2} + p^{2} + ({\bf k} + {\bf p})^{2})} \nonumber \\
&= \frac{3 G_{\epsilon}^{2}}{2 \epsilon}
\left[1 + \frac{\epsilon}{2} (3 - \ln 3) + O(\epsilon^{2})
\right] .
\end{eqnarray}
The finite parts of the integrals are required to calculate
the coefficient $K$ in the scaling function~(\ref{scalfun})
and the normalization constants $a$ and $b$~(\ref{a}, \ref{b}).

\end{document}